\documentclass[10pt,aps,twocolumn,showpacs,amsmath,amssymb,prx]{revtex4-1}
\usepackage{times}
\usepackage{stackengine}
\usepackage{graphicx}
\usepackage{color}
\usepackage{psfrag}
\usepackage{epsfig}
\usepackage[FIGTOPCAP]{subfigure}
\usepackage{bm}% bold math
\allowdisplaybreaks 
\usepackage{braket}
\usepackage{wasysym}
\usepackage[breaklinks]{hyperref}
\hypersetup{colorlinks=true, linkcolor=blue, citecolor=blue, filecolor=blue, urlcolor=blue}

\newcommand*\tlbl{\parbox[c][2.1em][c]{1.9em}{\includegraphics[width=1.5em,angle=90,origin=b]{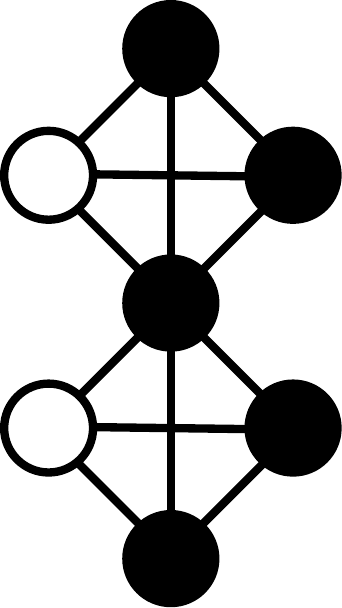}}}
\newcommand*\tlbleq{\parbox[c][2.1em][c]{1.6em}{\includegraphics[width=0.9em,angle=90,origin=b]{tlbl}}}
\newcommand*\tlbr{\parbox[c][2.1em][c]{1.9em}{\includegraphics[width=1.5em,angle=90,origin=b]{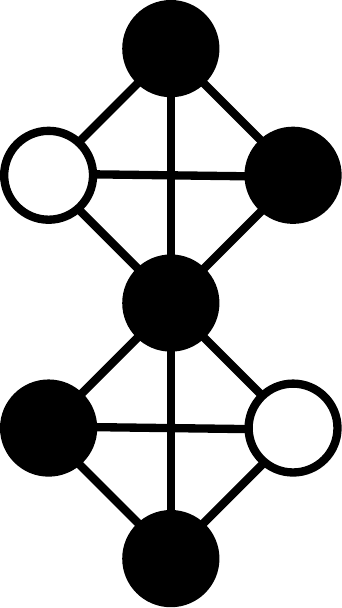}}}
\newcommand*\tlbreq{\parbox[c][2.1em][c]{1.6em}{\includegraphics[width=0.9em,angle=90,origin=b]{tlbr}}}
\newcommand*\tlbd{\parbox[c][2.1em][c]{1.9em}{\includegraphics[width=1.5em,angle=90,origin=b]{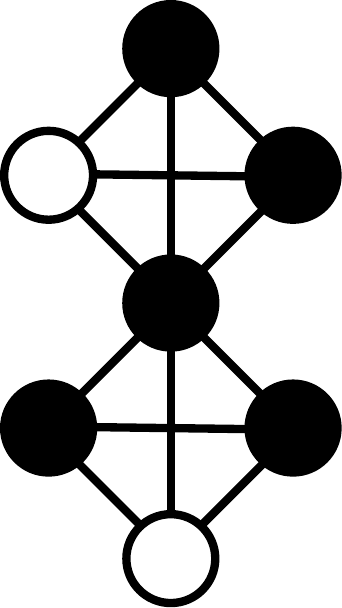}}}
\newcommand*\tlbdeq{\parbox[c][2.1em][c]{1.6em}{\includegraphics[width=0.9em,angle=90,origin=b]{tlbd}}}
\newcommand*\tubd{\parbox[c][2.1em][c]{1.9em}{\includegraphics[width=1.5em,angle=90,origin=b]{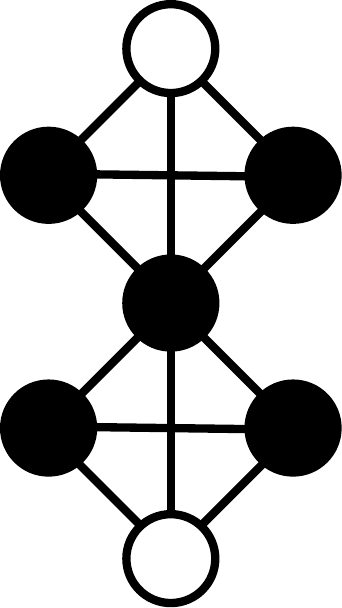}}}
\newcommand*\tubdeq{\parbox[c][2.1em][c]{1.6em}{\includegraphics[width=0.9em,angle=90,origin=b]{tubd}}}
\newcommand*\tdbu{\parbox[c][2.1em][c]{1.9em}{\includegraphics[width=1.5em,angle=90,origin=b]{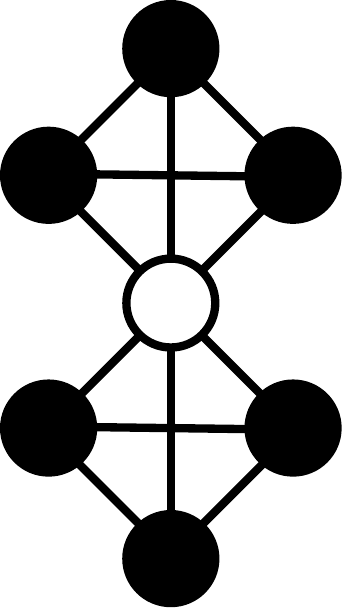}}}
\newcommand*\smon{\parbox[c][2.1em][c]{1.9em}{\includegraphics[width=1.5em,origin=b]{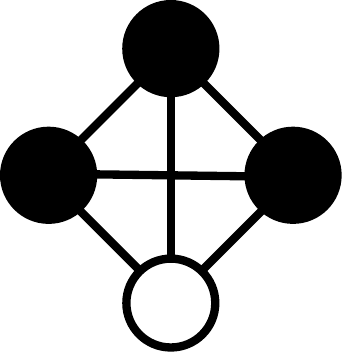}}}

\def\DTO{Dy$_2$Ti$_2$O$_7$}
\def\HTO{Ho$_2$Ti$_2$O$_7$}
\def\YTO{Yb$_2$Ti$_2$O$_7$}

\begin{document}

\title{Free coherent spinons in quantum square ice}

\author{Stefanos Kourtis}
\affiliation{Department of Physics, Princeton University, Princeton, NJ 08544, USA}
\author{Claudio Castelnovo}
\affiliation{TCM group, Cavendish Laboratory, University of Cambridge, Cambridge CB3 0HE, United Kingdom}

\date{\today}

\begin{abstract}
We investigate the quantum dynamics of monopole-like excitations in quantum square ice, as captured by the strongly anisotropic spin-1/2 XXZ model on the checkerboard lattice. We obtain exact results for excitation dynamics in both analytically solvable effective models and a fully interacting model of quantum square ice on finite clusters. We find that the dispersive lower bound of the dynamic response of freely propagating spinons is recovered in the dynamic structure factor of the interacting system, yielding a marked fingerprint of coherent spinon dispersion. Our results provide unbiased evidence for the formation of coherent quasiparticles propagating freely in the correlated ``vacuum'' of quantum square ice.
\end{abstract}

\maketitle

\section{Introduction and outline}

Frustrated quantum magnets are one of the most promising platforms for the realization of unconventional quantum liquids: states that elude characterization by local order parameters and host emergent quasiparticles carrying fractions of the microscopic degrees of freedom of the original system~\cite{Balents2010}. Such states often arise in quantum systems with disordered classical counterparts. Spin ice --- the manifold of extensively degenerate ground states of the pyrochlore Ising antiferromagnet --- is a remarkable example~\cite{Bramwell2001,Castelnovo2012,Castelnovo2015}. Discovered in materials that contain large magnetic moments of rare-earth ions, such as {\DTO} and {\HTO}, it is a classical spin liquid~\cite{Moessner1997} that maintains a residual entropy as the temperature tends towards zero~\cite{Ramirez1999} and supports magnetic monopole excitations~\cite{Ryzhkin2005,Castelnovo2008}. Materials isostructural to {\DTO} and {\HTO}, but with smaller moments and larger quantum fluctuations, have been identified as candidate \textit{quantum} spin ices: a highly sought-after quantum spin liquid state with fractional quasiparticles~\cite{Hermele2004,Banerjee2008,Gingras2014,Fennell2014}.  

Due to the prospect of realizing a quantum spin liquid, quantum spin ice and its excitations have been the target of intense theoretical and experimental activity~\cite{Ross2011,Savary2011,Shannon2012,Benton2012,Lee2012a,Savary2013,Hao2014,Applegate2012,Hayre2013,Kimura2013,Hirschberger2015a,Kato2015,Pan2015,Tokiwa2015}. However, finding observables that may be used as definitive signatures of quantum spin liquid behavior has proven a surprisingly tall order to date. Very recently, progress has been made in the context of other candidate spin liquid materials by evaluating dynamic responses~\cite{Knolle2015,Perreault2015}. 

Of particular and timely interest for quantum spin ice is the dynamics of its monopole excitations. Indeed, recent experimental efforts have claimed to observe behaviour consistent with coherently dispersing quantum monopoles in terahertz response~\cite{Pan2015} and thermal transport~\cite{Tokiwa2015}. Furthermore, recent inelastic neutron scattering experiments have revealed broad excitation continua in Yb-based compounds, which are expected to be proximate to quantum spin ice~\cite{Hallas2016}. Monopoles in classical spin ice are deconfined, which makes it plausible that their quantum counterparts delocalize and behave as coherent quasiparticles. Yet they move in and are in fact the product of a highly disordered and strongly correlated ``vacuum'', which may preempt coherent propagation via effective interactions. A number of recent theoretical efforts addressed the nature and dynamics of monopoles in quantum spin ice. For instance, it was shown that when quantum fluctuations are suppressed by the application of an external longitudinal field, monopole propagation can be effectively described by the free motion of quantum strings~\cite{Wan2012}. In zero field instead, toy models of monopoles propagating via single-spin flip processes exhibit spectra with a characteristically well-defined bandwidth~\cite{Petrova2015,Wan2015a}. All the aforementioned theoretical studies of excited quantum spin ice rely however on substantial approximations.

The main question we address in this work is: to which extent can monopoles be considered \textit{freely propagating coherent quasiparticles}? To answer, we obtain exact numerical results for the dynamics of monopoles in the two-dimensional analogue of quantum spin ice and compare them to effective free theories of monopole propagation.  Our interacting model and numerical method are a priori unbiased, contrary to the effective modeling found in the existing quantum spin ice literature~\cite{Ross2011,Savary2011,Lee2012a,Savary2013,Hao2014}. We discover signatures of coherent quantum propagation of free monopoles in the dynamic structure factor (DSF) obtained by exact diagonalization of the planar-pyrochlore XXZ model in the strongly anisotropic limit.

The DSF for the spin degrees of freedom derives from the correlations between creation and annihilation of nearest-neighbour pairs of monopoles at different points in space and time. One may thus wonder how much of the signal may be due to a monopole-antimonopole pair propagating as a bound object. We address this issue in detail by considering the additional scenario where the monopole pair is constrained to remain at nearest-neighbour distance (confined), where moreover substantial analytical progress can be made in understanding the behaviour of the DSF. By comparing the energetics in the confined case with that of deconfined excitations, we are able to ascertain that the lowest-energy dynamical processes are those where monopole and antimonopole propagate separately as free particles, even in the finite clusters we model numerically. Specifically, we find that a free monopole theory describes the dispersion of the bottom of the two-monopole excitation continuum observed in the DSF spectrum rather well, whereas higher-energy features indicate correlation-driven quasiparticle decoherence. Our results provide concrete evidence that the dynamic structure factor --- a quantity that can be accessed experimentally, e.g., via inelastic neutron scattering --- contains identifiable fingerprints of the so-far elusive behavior of fractionalized excitations in quantum spin ice.

Contrary to three dimensions, for the planar-pyrochlore (namely, checkerboard) antiferromagnet the consensus is that the ground-state phase diagram contains only ordered phases~\cite{Fouet2003,Tchernyshyov2003,Brenig2004,Starykh2005,Moukouri2008,Chan2011,Bishop2012}. However, taking into account virtual excitations, effective plaquette models~\cite{Moessner2004,Shannon2004,Syljuasen2006,Banerjee2013} inspired by dimer hamiltonians~\cite{Rokhsar1988} can be fine-tuned to a Rokhsar-Kivelson (RK) point by the addition of appropriate potential terms. The ground state at this point is a critical equal amplitude and phase superposition of all ice configurations --- a disordered spin liquid called \textit{quantum square ice}~\cite{Moessner2004,Shannon2004,Henry2014}. Recent quantum Monte Carlo results on quantum square ice have shown that an arbitrarily small density of violations of the ice rule, free to move in the otherwise frozen background, is enough to expand the critical point into a quantum $U(1)$ liquid phase~\cite{Henry2014}. Despite the differences between 2D and 3D, excited quantum square ice provides a viable playground to study the dynamics of fractional quasiparticles in quantum spin ice and to obtain some of its general characteristics. Moreover, even if not directly representative of a real material, square ice has in fact direct experimental implications. For example, classical square ice has been recently shown to describe graphene nanocapillaries~\cite{Algara-Siller2015}, whereas quantum square ice offers the interesting prospect to capture the behaviour of tailored ultracold atomic systems in optical lattices~\cite{Glaetzle2014}. 

\begin{figure}[t]
 \centering
 \subfigure{\topinset{(a)}{\includegraphics[width=0.4\columnwidth]{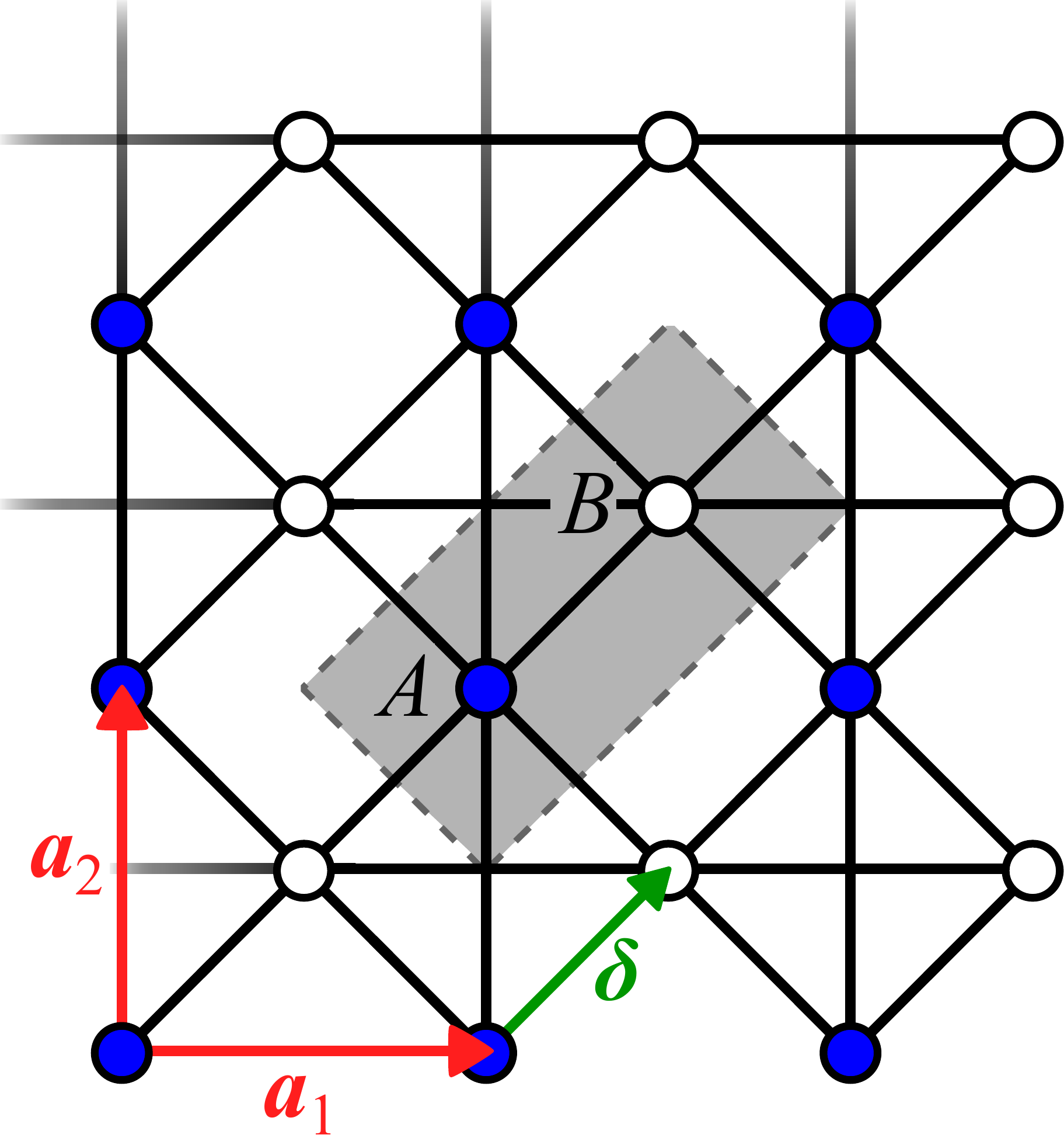}}{.0in}{-0.25\columnwidth}} \ \ \
 \subfigure{\topinset{(b)}{\includegraphics[width=0.45\columnwidth]{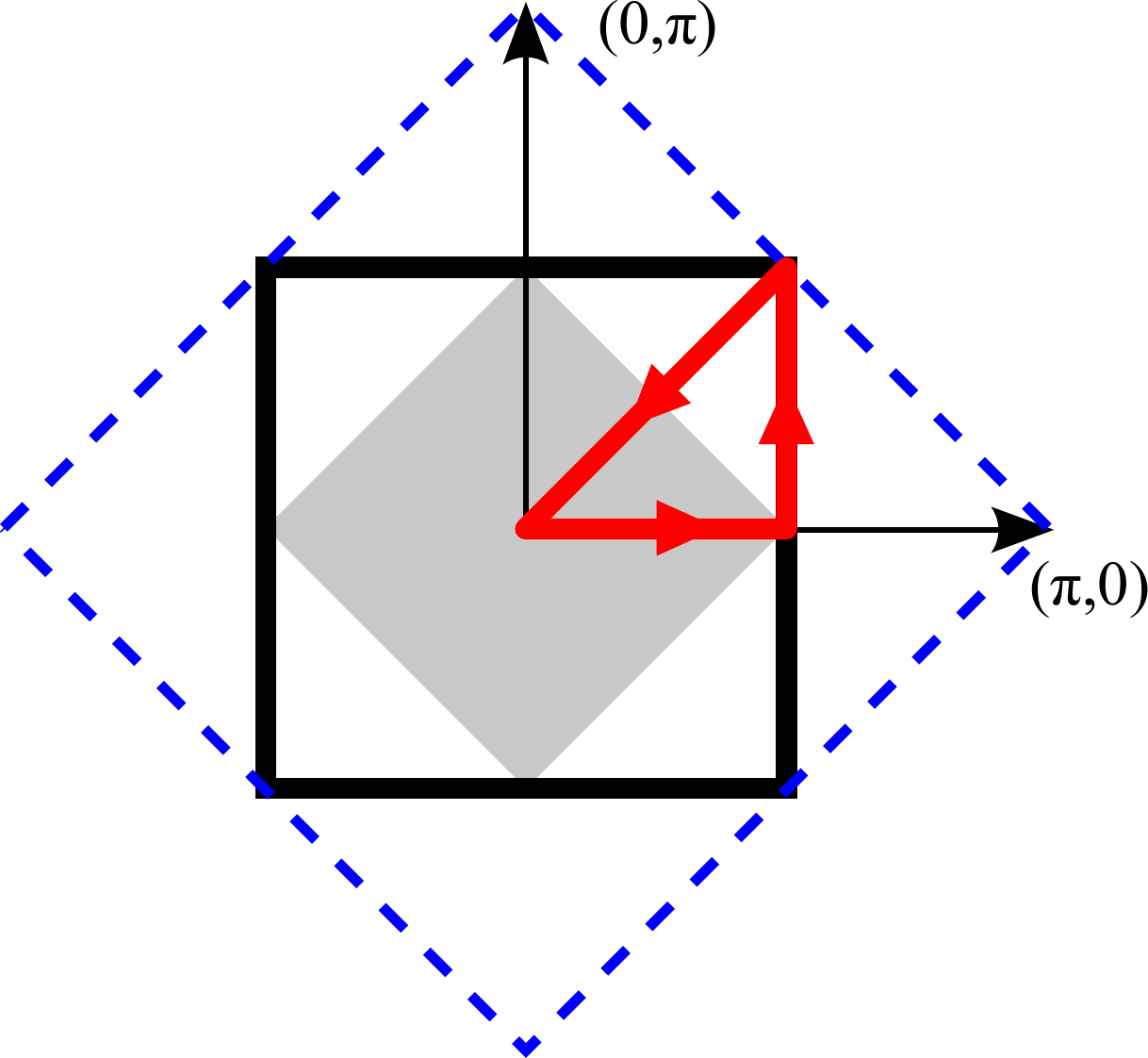}}{.0in}{-0.2\columnwidth}}
 \caption{(a) The checkerboard lattice $\Lambda$ with primitive translation vectors $\bm{a}_1 = (2 \,, 0)^\mathsf{T}$, $\bm{a}_2 = (0 \,, 2)^\mathsf{T}$ and basis vector $\boldsymbol\delta = (1 \,, 1)^\mathsf{T}$ connecting sublattices $\Lambda_A$ and $\Lambda_B$. (b) Brillouin zones of the checkerboard lattice (solid line) and the monopole lattice (shaded area). The red line indicates the high-symmetry path used below. The dashed line indicates the boundary of the compact support of the energy-resolved joint density of states.}
 \label{fig:ckb}
\end{figure}

Our exposition is organized as follows. In Sec.~\ref{sec:model}, we introduce the interacting model and necessary formalism. We then formulate effective free theories to describe: (a) confined motion, in which monopoles are artificially constrained to be at nearest-neighbor distance; and (b) the unconstrained deconfined case pertinent to quantum square ice. In the confined case, we identify two distinct microscopic processes which allow for an analytical understanding of the behaviour of the system in terms of self-avoiding polygons and Lieb chains. We present our main numerical results and the comparison to free monopole theory in Sec.~\ref{sec:results} and our conclusions in Sec.~\ref{sec:conclusions}. For convenience and consistency, we have so far used the term ``monopole'' for spinon excitations in both 3D and 2D spin ice systems. In the rest of the manuscript we shall refer to such excitations in square ice as spinons, except where calling them monopoles and antimonopoles appeals more directly to intuition.

\section{Microscopic model and effective theories}\label{sec:model}

\subsection{The checkerboard-lattice XXZ model}

Consider a checkerboard lattice $\Lambda$ formed by two sublattices $\Lambda_A$ and $\Lambda_B$, as shown in Fig.~\ref{fig:ckb}(a). We define the XXZ model on $\Lambda$ as
\begin{equation}
 {\cal H}_{\mathrm{XXZ}} = \sum_{\langle \bm{i}, \bm{j} \rangle} J_z S^z_{\bm{i}} S^z_{\bm{j}} - \frac{J_\pm}{2} ( S^+_{\bm{i}} S^-_{\bm{j}} + S^-_{\bm{i}} S^+_{\bm{j}} ) \,, \label{eq:xxz}
\end{equation}
where $S^x_{\bm{i}}, \, S^y_{\bm{i}},\, S^z_{\bm{i}}$ are the projections of a spin-1/2 degree of freedom on site $\bm{i}\in\Lambda$, $S_{\bm{i}}^\pm = S_{\bm{i}}^x \pm \mathrm{i} S_{\bm{i}}^y$, and the real parameters $J_z,J_\pm>0$ are the strengths of the Ising and transverse interactions, respectively. The shorthand $\langle  \bm{i} , \bm{j} \rangle$ denotes that $\bm{j}$ is nearest neighbor of $\bm{i}$ (in the checkerboard lattice sense), as shown in Fig.~\ref{fig:ckb}(a). In the classical regime $J_\pm = 0$, the frustration of the Ising interaction leads to an extensively degenerate manifold of ground states that obey the so-called ice rule~\cite{Nagle1966,Lieb1967a,Lieb1967}. This can be expressed in various equivalent ways, such as the ``2-in -- 2-out'' constraint or vanishing magnetization on every crossed plaquette.

\begin{figure}[t]
 \centering
 \includegraphics[width=0.9\columnwidth]{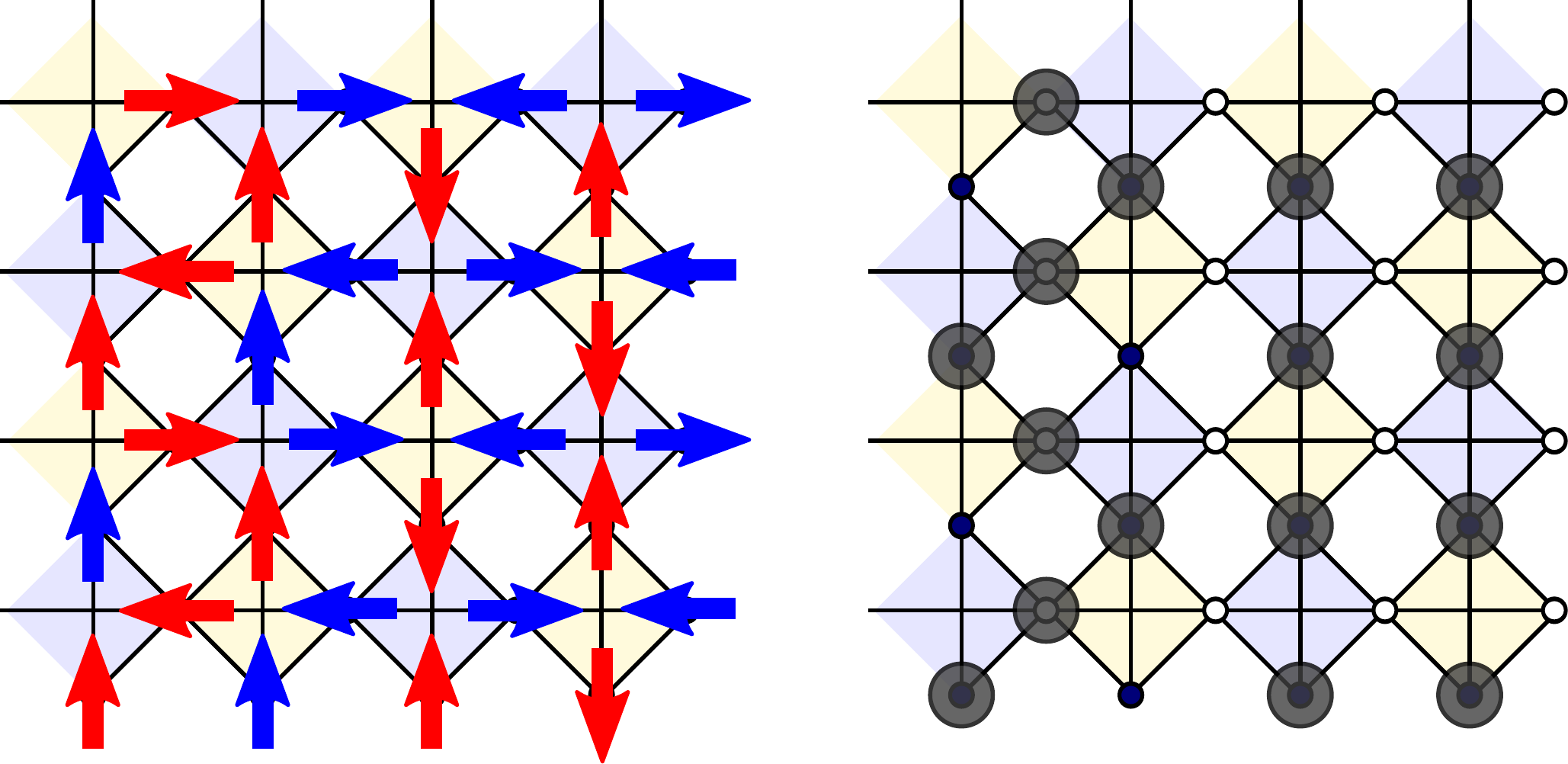}
 \caption{Example of a spin ice configuration and its translation to the hardcore boson language. Red (blue) arrows represent $S^z_{\bm{i}}=+1/2$ ($S^z_{\bm{i}}=-1/2$). Blue and yellow crossed plaquettes define the two spinon sublattices, defined in the main text. Red and blue arrows are used to visualize the ice rule on each of the two monopole sublattices: blue (red) spins are ``out'' in the blue (yellow) plaquettes, whereas red (blue) spins are ``in'' in the blue (yellow) plaquettes. The configuration shown here contains a stripe of antiferromagnetically ordered spins, pertinent to the discussion of Sec.~\ref{subsec:confined}.}
\label{fig:spinboson}
\end{figure}

The XXZ model can be recast into a model of interacting hardcore bosons via the mapping $S^+_{\bm{i}} \longleftrightarrow a_{\bm{i}}^\dagger$, $S^z_{\bm{i}} \longleftrightarrow 1/2 - n_{\bm{i}}$, where $a_{\bm{i}}$ ($a_{\bm{i}}^\dagger$) annihilates (creates) a boson at position $\bm{i}$ and $n_{\bm{i}}=a_{\bm{i}}^\dagger a_{\bm{i}}$. The bosonic hamiltonian
\begin{equation}
 {\cal H}_{\mathrm{XXZ}} = \sum_{\langle \bm{i}, \bm{j} \rangle} J_z \left( n_{\bm{i}} - \frac12 \right) \left( n_{\bm{j}} - \frac12 \right) - \frac{J_\pm}{2} \left( a_{\bm{i}}^\dagger a_{\bm{j}} + a_{\bm{j}}^\dagger a_{\bm{i}} \right) 
%\,, 
\label{eq:boson}
\end{equation}
is fully equivalent to the original XXZ model of Eq.~\eqref{eq:xxz}. The real parameters $J_\pm$ and $J_z$ now correspond to the hopping matrix element and the nearest-neighbor interaction strength, respectively. In the boson language, the ice rule is obeyed by configurations with precisely two bosons on every crossed plaquette, and all spin-ice ground states are at half filling. Equivalently, the ice rule can be stated as constraining each boson to having precisely two bosons neighboring it. The latter formulation can be pictured as bosons forming closed loops in spin-ice states. Even though here we adopt the bosonic representation of the XXZ model mainly for illustrative purposes, its physicality has been recently put forward in connection to ultracold atoms in optical lattices~\cite{Glaetzle2014}.

In this work, we concern ourselves with the hardcore-Ising limit $J_z = \infty$, restricting the system to the degenerate manifold $M_0$ of low-energy many-body configurations $\ket{c}$ obeying the ice rules. The subspace $M_0$ can be formally defined as
\begin{equation}
 M_0 = \left\lbrace \ket{c} \ : \ \braket{c| \sum_{\bm{j} \in \XBox} n_{\bm{j}} |c} = 2, \; \forall \, \XBox \in \Lambda \right\rbrace \, ,
\end{equation}
where the sum is over all sites in one of the crossed plaquettes of the checkerboard lattice denoted by $\XBox$. The states in $M_0$ are not directly connected by the transverse term $J_\pm$. Nevertheless, we choose the initial state to be the equal amplitude and phase superposition of the states in $M_0$. We denote this state by $\ket{\mathrm{RK}}$, because it is the ground state at the Rokhsar-Kivelson (RK) point of the plaquette-flip model of Refs.~\cite{Shannon2004,Moessner2004} --- the only point in the phase diagram of the system where it is in a (critical) liquid state, which is the focus of our work. Notice that this is also an eigenstate of ${\cal H}$, since ${\cal H}_{\mathrm{XXZ}}\ket{\mathrm{RK}} = 0$ in the $J_z \rightarrow \infty$ limit. 

Specifically, we are interested in the dynamics of a single spin-flip introduced in the otherwise perfect ice-rule-satisfying $\ket{\mathrm{RK}}$ state. This corresponds to doping the half-filled system with one particle or hole in the boson language. Bosons can now hop between states for which the ice rule is violated at precisely two crossed plaquettes, thus restoring kinetics in the system. Equivalently, a single spin flip introduces two violations of the ``2-in -- 2-out'' ice rules, namely a ``3-in -- 1-out'' and a ``3-out -- 1-in'' plaquette in the checkerboard lattice. These defects can be viewed as opposite gauge charges and correspond to the magnetic monopoles in (pyrochlore) spin ice~\cite{Castelnovo2012}.

\begin{figure}[t]
 \centering
 \includegraphics[width=0.8\columnwidth]{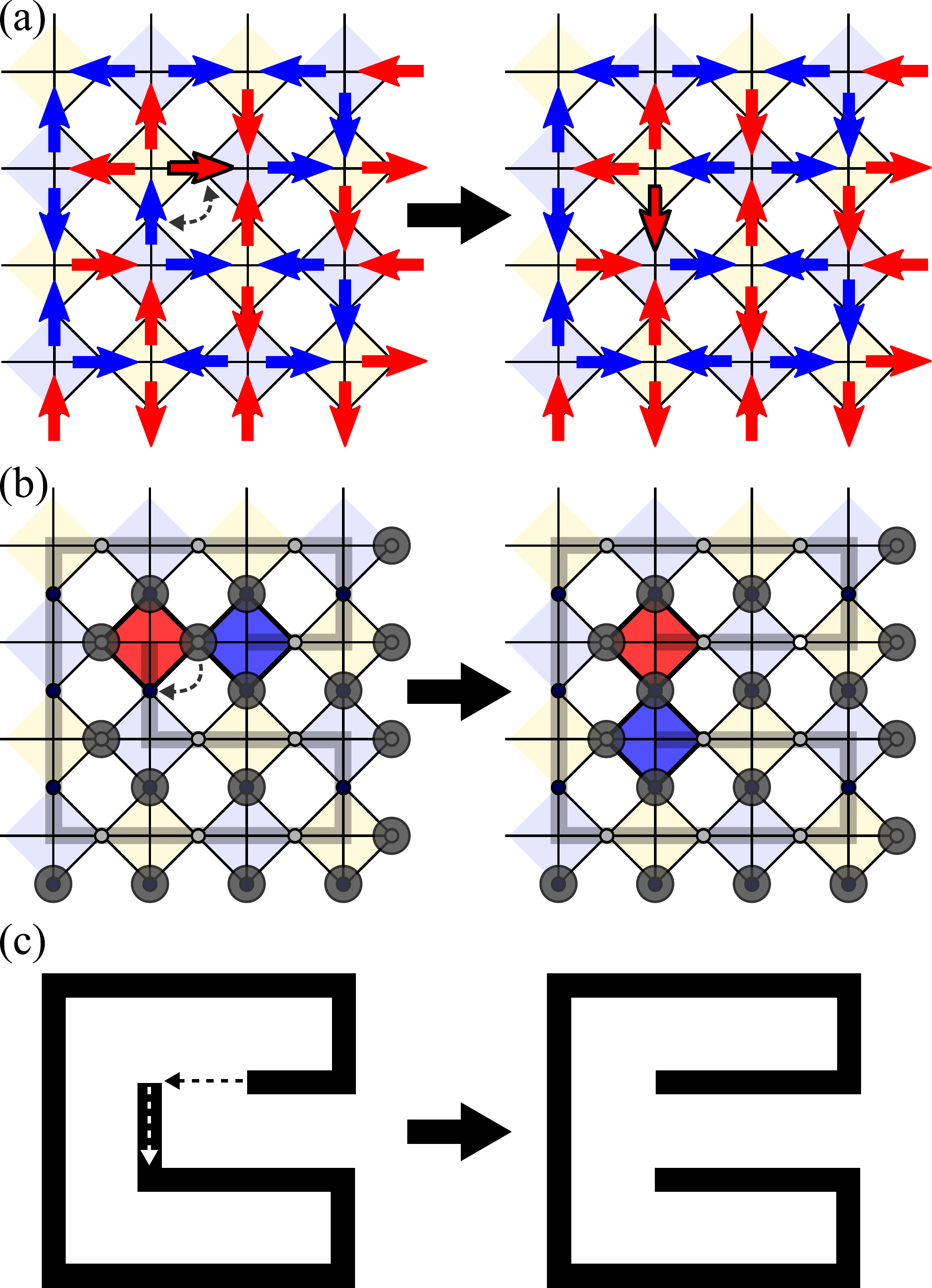}
 \caption{Illustration of the leapfrog process in (a) the spin language, (b) the boson language, and (c) the self-avoiding polygon language. The leapfrogging dipole cannot deviate from the ``track'' laid by the spin background.}
 \label{fig:leapfrog}
\end{figure}

Without loss of generality, in the boson representation we choose to dope the system with an extra particle and denote the corresponding two-spinon manifold by $M_{+2}$. Restricted to act within $M_{+2}$ to lowest (zeroth) order in $J_z$, the hamiltonian becomes the single term
\begin{subequations}
\begin{equation}
 {\cal H} = - J_\pm \sum_{\langle \bm{i},\bm{j} \rangle} \left( a_{\bm{i}}^\dagger F_{\bm{ij}} a_{\bm{j}} + \mathrm{h.c.} \right) \,,
\end{equation}
where
\begin{equation}
 F_{\bm{ij}} = \left( 2 - \sum_{\langle \bm{m},\bm{i} \rangle} n_{\bm{m}} \right) \left( 2 - \sum_{\langle \bm{l},\bm{j} \rangle} n_{\bm{l}} \right) \,.
\end{equation}\label{eq:model}%
\end{subequations}
The projector $F_{\bm{ij}}$ ensures that the quantum hopping processes conserve spinon number. This can be seen as follows. Within $M_{+2}$, the action of $\left( 2 - \sum_{\langle \bm{l},\bm{j} \rangle} n_{\bm{l}} \right) a_{\bm{j}}$ on any state yields a nonzero result only if one of the two plaquettes containing site $\bm{j}$ is triply occupied. The projector thus ensures that $a_{\bm{j}}$ acts only on plaquettes occupied by monopoles. Similarly, $a_{\bm{i}}^\dagger \left( 2 - \sum_{\langle \bm{m},\bm{i} \rangle} n_{\bm{m}} \right)$ is nonzero only if $\bm{i}$ belongs to plaquettes that obey the ice rule. The above two constraints combined ensure that, overall, no monopole creation or annihilation events are allowed. Even though in this work we shall focus exclusively on $M_{+2}$, this hamiltonian can be extended to any $2q$-spinon sector, $q \in \mathbb{N}$. Notice that the processes described by Eq.~\eqref{eq:model} preserve the ``spinon sublattice'' (depicted as blue / yellow plaquettes in Fig.~\ref{fig:spinboson} above), i.e., spinon hopping occurs only between next-nearest neighboring plaquettes --- see examples in Figs.~\ref{fig:leapfrog} and~\ref{fig:noleapfrog}. Spinons of opposite charge therefore sit strictly on different sublattices, which straightforwardly forbids annihilation processes. More details on the properties of ${\cal H}$ are provided in App.~\ref{sec:rk}.

By obtaining the eigenpairs of ${\cal H}$ in the two-spinon sector, we are then able to evaluate the dynamic structure factor (DSF): 
\begin{subequations}
\begin{align}
 {\cal S}(\bm{k},\omega) =&{\ } \sum_{s=A,B} \sum_m | \braket{m| S_{s,\bm{k}}^+ |\mathrm{RK}} |^2 \delta( \omega - E_m ) \\
 =&{\ } \sum_{s=A,B} \sum_m | \braket{m| a_{s,\bm{k}}^\dagger |\mathrm{RK}} |^2 \delta( \omega - E_m ) \,,
\end{align}\label{eq:dsf}%
\end{subequations}
where $\ket{m}$ stands for the eigenstates of ${\cal H}$ in $M_{+2}$ with energy $E_m$ and $S_{s,\bm{k}}^+ = \sum_{\bm{i}\in\Lambda_s} e^{\mathrm{i}\bm{k}\cdot\bm{i}} S_{\bm{i}}^+$, $a_{s,\bm{k}}^\dagger = \sum_{\bm{i}\in\Lambda_s} e^{\mathrm{i}\bm{k}\cdot\bm{i}} a_{\bm{i}}^\dagger$. Note that all our results are the same irrespective of whether we evaluate the DSF with $S_{s,\bm{k}}^+$ or $S_{s,\bm{k}}^-$. In the following, we will use $S_{s,\bm{k}}^+$, i.e., $a_{s,\bm{k}}^\dagger$, and correspondingly focus on $M_{+2}$.

\subsection{Effective theories of spinon dynamics} 

Spinons confined to nearest-neighbor distance can move via two distinct processes under the action of ${\cal H}$: one flips their common spin and one does not. It is interesting to study the two processes separately. 

In order to understand the two-spinon hopping dynamics, it is convenient to distinguish between the case of the two spinons remaining at nearest-neighbour distance, and the case where they move apart from one another. Being deconfined, we expect their dynamics to be a combination of the two processes; however, we can artificially constrain the two spinons to move together (which we refer to as ``confined propagation'' below), in which case they behave effectively as a single (dipolar) hopping particle. This allows for a higher degree of analytical understanding, which then proves useful in understanding the behaviour of the system by contrasting it with the far more challenging generic case of deconfined propagation. 

\subsubsection{Confined propagation}\label{subsec:confined}

The first process was dubbed ``leapfrog'' in the literature~\cite{Shannon2004}
and the corresponding hamiltonian can be written as 
\begin{equation}
 {\cal H}_{\mathrm{leapfrog}} = - J_\pm \sum_{\langle \bm{i},\bm{j} \rangle} ( a_{\bm{i}}^\dagger P_{M_0} a_{\bm{j}} + \mathrm{h.c.} ) \,,
\end{equation}
where $P_{M_0}$ is a projector that enforces no violations of the ice rules, i.e., a projector onto the spin-ice manifold $M_0$. When motion is constrained to leapfrog only, each confined configuration in the two-spinon sector contains a ``rail'' on which the dipole can move~\cite{Jaubert2011}. This is illustrated in Fig.~\ref{fig:leapfrog}. Indeed, the repetition of leapfrog moves flips each spin involved twice, except for the starting and ending spins; as a result, winding a spinon pair around a closed track back to its original position does not change the spin configuration at all. Since the motion on a track can only be backwards or forwards, each many-body state is connected to precisely two other states. Applying the Perron-Frobenius theorem to the hamiltonian matrix in this regime immediately yields the ground-state energy $E_0/t=-2$ and the bandwidth $W/t=4$. Note that, since it cannot modify the spins outside a given path, ${\cal H}_{\mathrm{leapfrog}}$ is necessarily non-ergodic.

\begin{figure}[t]
  \centering
 \includegraphics[width=\columnwidth]{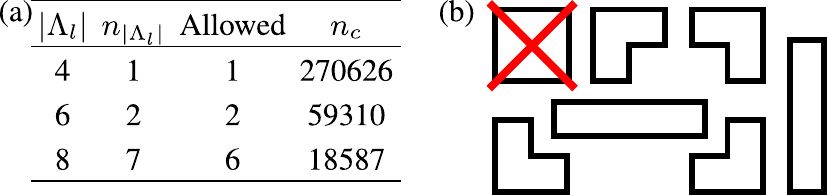}
  \caption{(a) Table of self-avoiding polygons (SAPs): length $|\Lambda_l|$; total number of SAPs $n_{|\Lambda_l|}$ of length $|\Lambda_l|$; number of SAPs allowed by the ice rules; and number of configurations $n_c$ compatible with each SAP embedded into the 72-site cluster (modulo translations). (b) The 7 SAPs of length 8; the SAP forbidden by the ice rules is crossed out.}%
  \label{fig:saps}%
\end{figure}

The interaction of the propagating dipole with the spin-ice background in the case of leapfrog is, in a sense, trivial: each configuration contains a track of spins, which are the only ones that are affected by the motion of the dipole. The problem therefore reduces to one-dimensional hopping on this track and the total dynamics is given by averaging over all possible tracks for a given lattice. These tracks are self-avoiding polygons (SAPs)~\cite{Guttmann2001} and their enumeration is a well-known problem in combinatorics, that has been performed algorithmically for the square lattice~\cite{Enting1980}. It is not necessary to obtain the extensive number of all SAPs for our purposes. The salient dynamical features of ${\cal H}_\mathrm{leapfrog}$ can be obtained by considering only SAPs of the shortest few lengths. Indeed, the probability of encountering a certain fixed path of spins decreases exponentially as the length of the path is increased. This is because of the concomitant exponential decrease of the number of spin-ice configurations on the complement of the path in the full lattice --- see Fig.~\ref{fig:saps}(a).

\begin{figure}[t]
 \centering
 \includegraphics[width=\columnwidth]{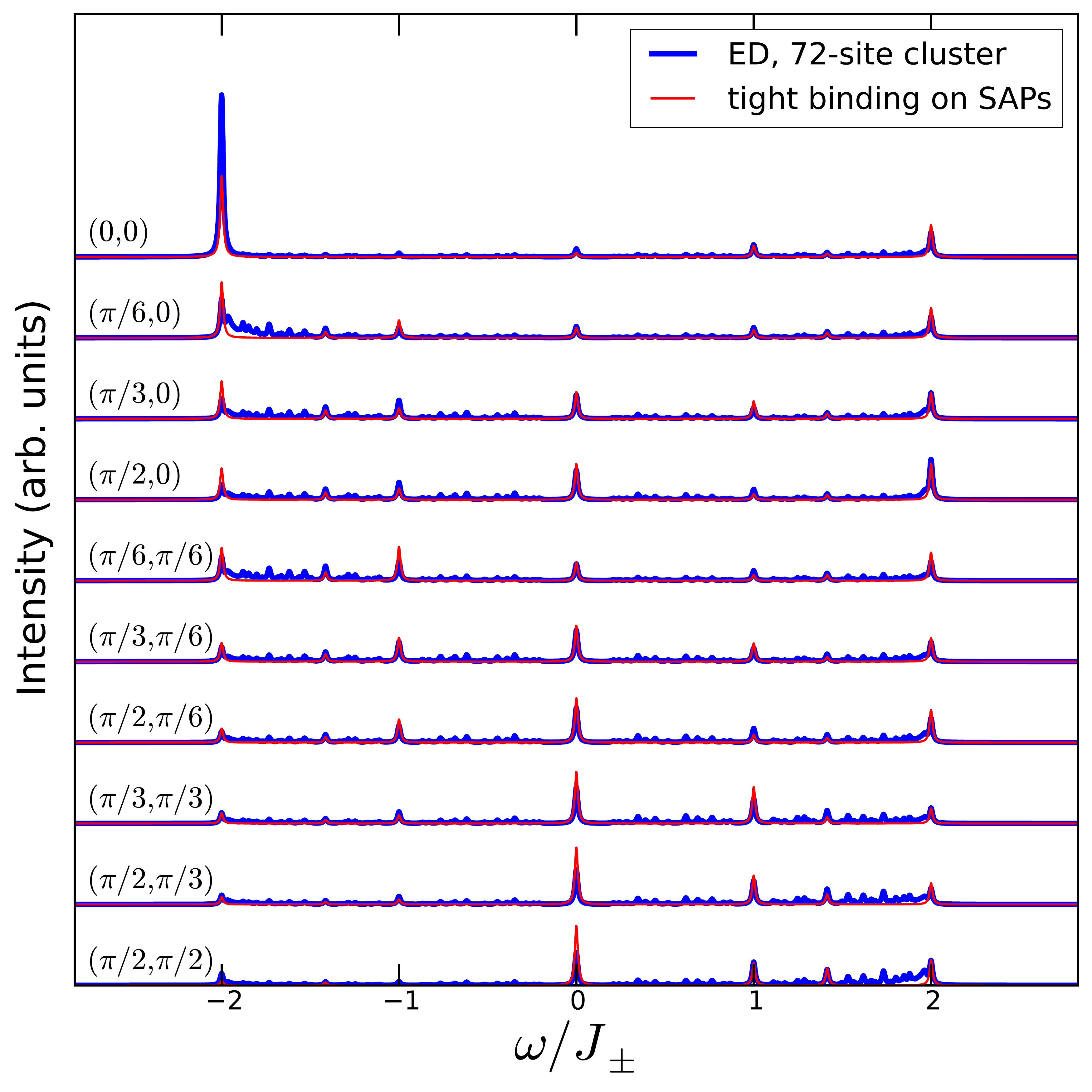}
 \caption{Comparison between the DSF obtained by ED of the leapfrog hamiltonian on a 72-site cluster and the dipole spectral function, obtained by combining the spectral function of tight binding models on SAPs of lengths 4, 6 and 8. All the momenta in the irreducible wedge of the Brillouin zone are shown.}
 \label{fig:akw-leapfrog}
\end{figure}

To confirm this, we evaluate the spectral function of the free dipole hopping on SAPs of lengths up to 8 steps. Consider a spin-ice configuration $\ket{c}$. The introduction of a nearest-neighbour monopole-antimonopole pair, $a_{\bm{i}}^\dagger \ket{c}$, uniquely identifies a SAP $\Lambda_l$ of length $|\Lambda_l|$, where the index $l$ runs over all SAPs. The only operators relevant to ${\cal H}_{\mathrm{leapfrog}}$ are those defined on $\Lambda_l$, effectively reducing it to a hopping hamiltonian on the closed path $\Lambda_l$. Note that, since dipoles cannot hop out of this closed path, $\Lambda_l$ can be effectively seen as a one-dimensional periodic lattice. It is therefore convenient to rewrite 
\begin{equation}
 a_{\bm{i}} = \frac{1}{|\Lambda_l|} \sum_{\kappa \in \Lambda_l^{-1}} e^{\mathrm{i} \kappa x_{\bm{i}}} a_{\kappa} \,,
\end{equation}
where $\kappa$ is a one-dimensional momentum belonging to the reciprocal lattice $\Lambda_l^{-1}$ and $x_{\bm{i}}$ is the position of site $\bm{i}$ within $\Lambda_l$. The matrix elements involved in the evaluation of the DSF can then be written as
\begin{equation}
\braket{m| a_{A\bm{k}}^\dagger |\mathrm{RK}} = \sum_{c} \sum_{\bm{i} \in \Lambda_A} \sum_{\kappa \in \Lambda_l^{-1}} \frac{e^{\mathrm{i} (\bm{k} \cdot \bm{i} - \kappa x_{\bm{i}}) }}{\sqrt{N_c}|\Lambda_l|} \braket{m| a_\kappa^\dagger |c} \,,
\label{eq:akw-leap}
\end{equation}
where $N_c$ is the number of spin-ice configurations $\ket{c}$ and $|\Lambda_l|$ the number of sites in $\Lambda_l$. Note that the contribution of each $\Lambda_l$ is weighted by the number of configurations containing the corresponding SAP, which can be obtained numerically for finite lattice sizes. We remark that some SAPs are incompatible with the ice rules, and the number of $\Lambda_l$ with fixed $|\Lambda_l|$ differs from the enumeration found in the literature~\cite{Enting1980}  --- see Fig.~\ref{fig:saps}(b).

We compare the (averaged) spectral function for tight binding models on SAPs with $\vert \Lambda_l \vert \leq 8$ to the full DSF of ${\cal H}_{\mathrm{leapfrog}}$ obtained numerically. The comparison is shown in Fig.~\ref{fig:akw-leapfrog}. We find very good agreement for all major features in the DSF and their dispersion, demonstrating that the dipole behaves as a \textit{coherently dispersing} particle, thanks to the quantum fluctuations produced by $J_\pm$. Minor discrepancies due to neglecting longer SAPs are expected.

\begin{figure}[t]
 \centering
 \subfigure{\topinset{(a)}{\includegraphics[width=0.48\columnwidth]{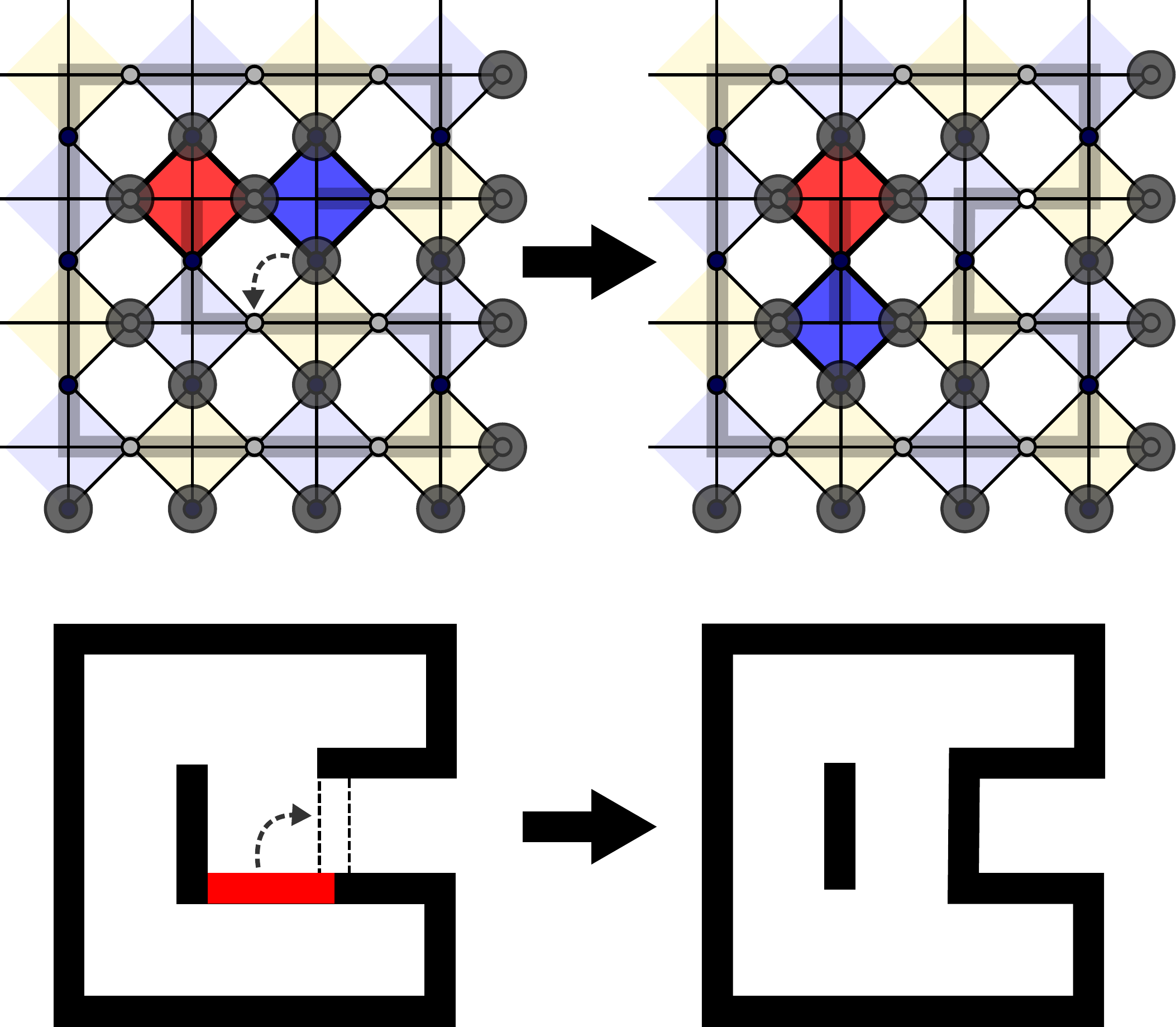}}{-.1in}{-0.8in}} \ \ \
 \subfigure{\topinset{(b)}{\includegraphics[width=0.48\columnwidth]{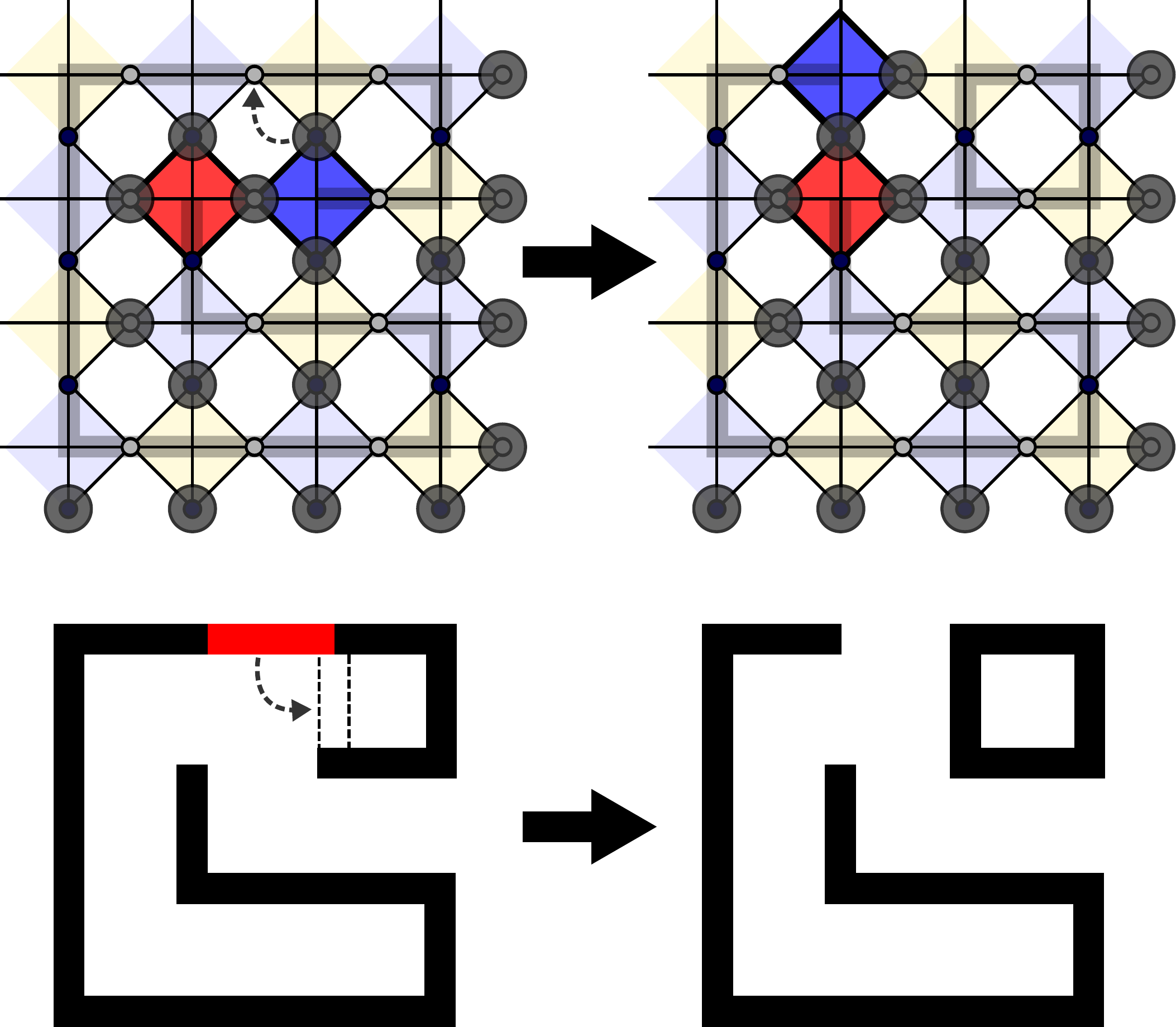}}{-.1in}{-0.8in}}
 \subfigure{\topinset{(c)}{\includegraphics[width=0.99\columnwidth]{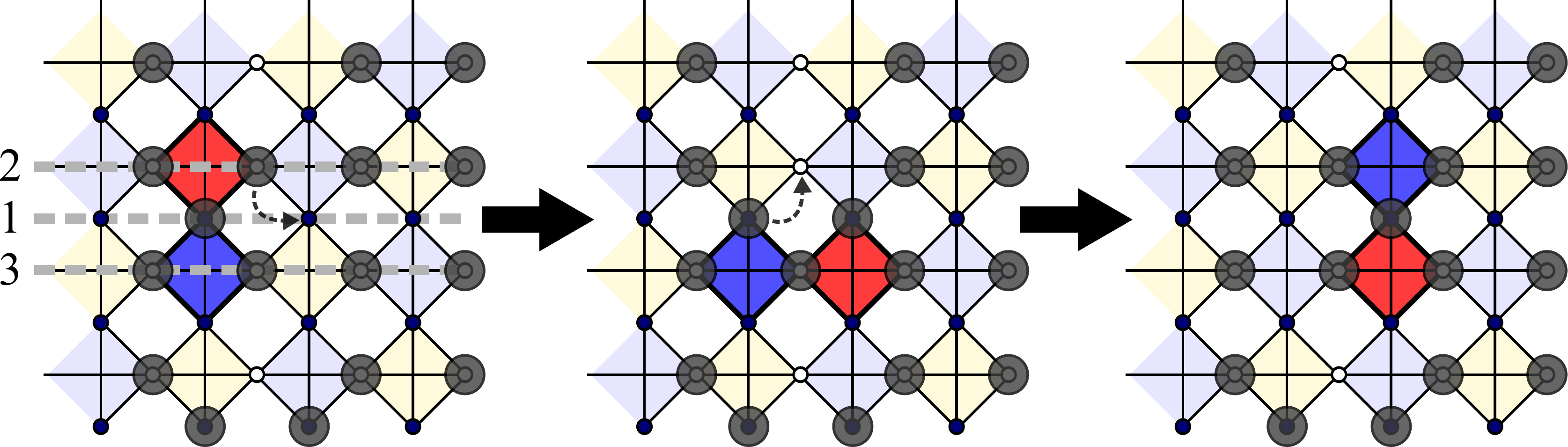}}{.0in}{-0.5\columnwidth}}
 \caption{(a,b) Illustration of reconnection processes and the underlying self-avoiding polygons. (c) Demonstration of dipole propagation on emergent one-dimensional Lieb chain; thick dashed lines follow the three sublattices of the 1D Lieb lattice.}
 \label{fig:noleapfrog}
\end{figure}

As mentioned above, nearest-neighbour spinons can move whilst remaining adjacent to one another also via a different process, which does not involve flipping the common spin between the spinons. We call this second process ``reconnection''. It is illustrated in Fig.~\ref{fig:noleapfrog} and is represented by the hamiltonian term 
\begin{equation}
 {\cal H}_{\mathrm{reconnection}} = - J_\pm \sum_{\langle \bm{i},\bm{j} \rangle} ( P_{\mathrm{nn}} \, a_{\bm{i}}^\dagger \left( 1 - P_{M_0} \right) a_{\bm{j}} P_{\mathrm{nn}} + \mathrm{h.c.} ) \,,
\end{equation}
where $P_{\mathrm{nn}}$ enforces the nearest-neighbor condition between monopole and antimonopole. Reconnection introduces nontrivial correlations of the dipole with its spin background. Contrary to the leapfrog process, the connectivity of a state via reconnection varies between 2 and 4, depending on the local spin arrangement. This irregularity in the connectivity of the Fock space means that we cannot expect reconnection to lead to a behaviour close to that of a free particle. 

Like leapfrog, reconnection is non-ergodic, and connected components of the Fock space with the most spins left unrestricted contribute the most to the dynamic response. With this reasoning, it is seen that the most significant processes arise in configurations that contain an antiferromagnetic stripe of small length. Such configurations, examples of which are shown in Figs.~\ref{fig:spinboson} and ~\ref{fig:noleapfrog}, give rise to dipole motion equivalent to hopping on finite one-dimensional Lieb lattices~\cite{Thouless1984,Lieb1989,Tasaki1992,Mielke1999}. We therefore anticipate signatures of the emergent one-dimensional dynamics on a Lieb chain to be identifiable in the overall reconnection spectral function, even though in this case the spectrum cannot be reduced to noninteracting contributions only.

\begin{figure}[t]
 \centering
 \includegraphics[width=\columnwidth]{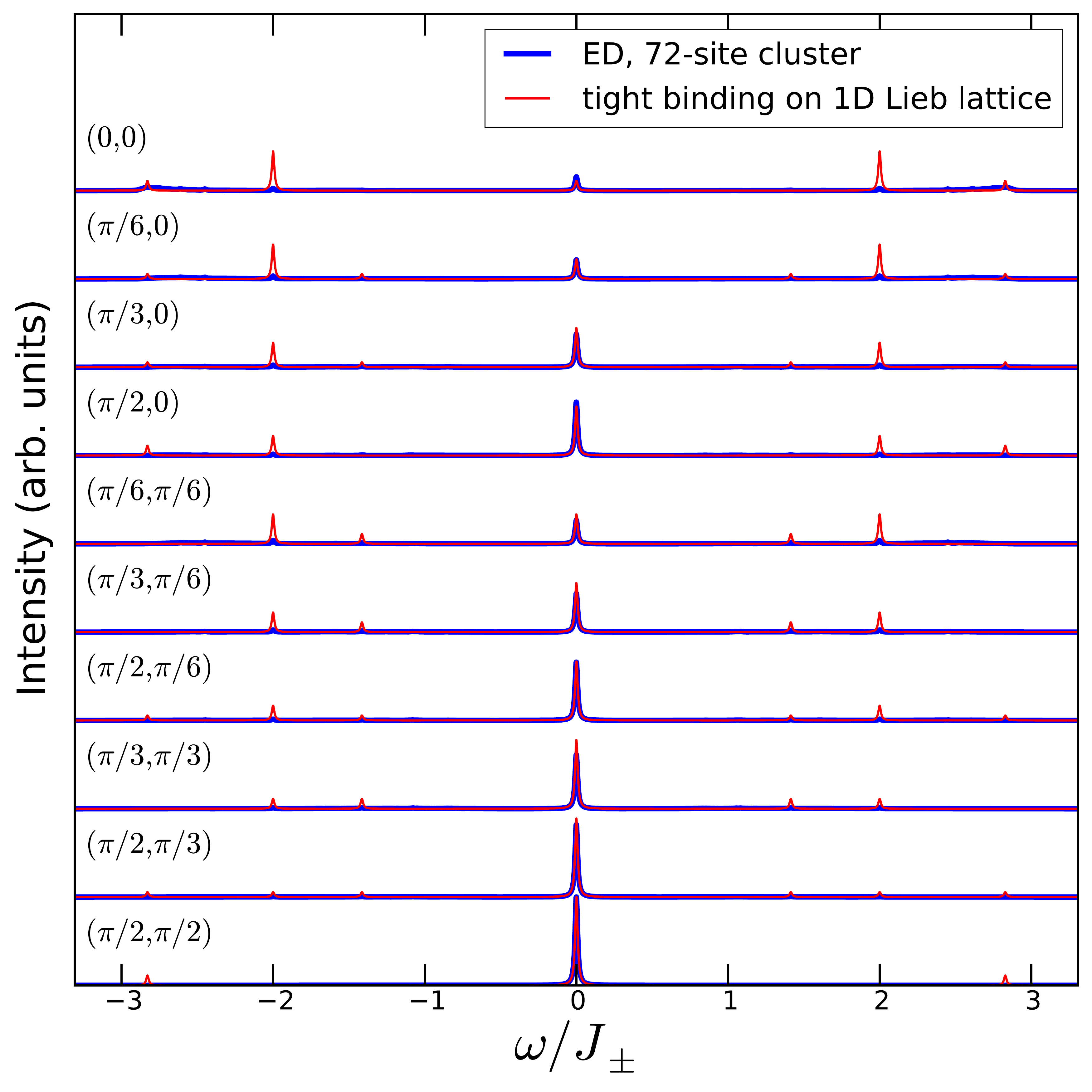}
 \caption{Comparison between the DSF obtained by ED of the reconnection hamiltonian on a 72-site cluster and the dipole spectral function obtained by combining the spectral function of tight binding on 1D Lieb chains along $x$ and $y$.}
 \label{fig:akw-noleapfrog}
\end{figure}

In analogy with the study of ${\cal H}_{\mathrm{leapfrog}}$, we make an explicit comparison between Lieb chains of up to 6 unit cells, both open and periodic wrapping around the lattice, and the exact calculation of the DSF of ${\cal H}_{\mathrm{reconnection}}$ in Fig.~\ref{fig:akw-noleapfrog}. The most prominent characteristic of Lieb lattices is a perfectly flat band at zero energy. This band is recovered in the fully interacting calculations of the spectral function of ${\cal H}_{\mathrm{reconnection}}$. The corresponding feature in the DSF, as well as the modulation of its intensity with momentum, are captured very well by the Lieb-chain picture. Features away from $\omega/J_\pm=0$ match corresponding features in the DSF in energy, but not in intensity. Nevertheless, the overall bandwidth of the interacting calculations matches precisely the $4\sqrt{2} J_\pm$ bandwidth of the 1D Lieb lattice. The inclusion of further noninteracting processes improves the agreement, but, as already mentioned, we do not expect a perfect match in this case. As a side note, we remark that whether the reconnection process is possible or not depends on the geometry of the lattice. For example, even though it is possible in the planar pyrochlore, it is not on the three-dimensional pyrochlore lattice.

Having identified the two relevant processes in the dynamics of a confined pair of spinons, we now take both into account on the same footing, 
\begin{equation}
 {\cal H}_{\mathrm{confined}} = {\cal H}_{\mathrm{leapfrog}} + {\cal H}_{\mathrm{reconnection}} \,.
\end{equation}
Reconnection can then be understood as a rearrangement of the SAP that the dipole leapfrogs on, which thereby destroys --- at least partially --- the coherent propagation on SAPs. Conversely, leapfrog allows the dipole to escape the Lieb chains formed in configurations like those in Fig.~\ref{fig:noleapfrog}(c). We note that, even though both ${\cal H}_{\mathrm{leapfrog}}$ and ${\cal H}_{\mathrm{reconnection}}$ are not ergodic on their own, their combination restores ergodicity. We have checked this explicitly for the 32-site cluster, using the procedure outlined in App.~\ref{sec:methods}.

\begin{table}[t]
  \centering
   \begin{tabular*}{\columnwidth}{ccccccccccc}
    \hline
    Configuration & Multiplicity & Amplitude & $\nearrow$ & $\searrow$ & $\swarrow$ & $\nwarrow$ & $\rightarrow$ & $\leftarrow$ & $\uparrow$ & $\downarrow$ \\
    \hline
    1. \tdbu & 1  & 0.1027 & 0.8 & 0.8 & 0.8 & 0.8 & 0 & 0 & 0 & 0 \\
    2. \tubd & 1  & 0.1293 & 0.8 & 0.8 & 0.8 & 0.8 & $\bm{1}$ & $\bm{1}$ & 0 & 0 \\
    3. \tlbl & 2  & 0.2366 & 0.8 & $\bm{1}$   & $\bm{1}$   & 0.8 & 0 & 0 & 0 & 0 \\
    4. \tlbr & 2  & 0.1387 & $\bm{1}$.5 & 0   & $\bm{1}$.5 & 0   & 0 & 0 & 0 & 0 \\
    5. \tlbd & 4  & 0.3927 & 0.8 & 0   & $\bm{1}$.5 & 0.8 & $\bm{1}$ & 0 & 0 & 0 \\
    \hline
  \end{tabular*}
  \caption{Possible configurations of an $x$-aligned dipole, their multiplicities under rotations and reflections, amplitudes in $a_{\bm{k}}^\dagger \ket{\mathrm{RK}}$, and cumulative dipole hopping amplitudes (in units of $J_\pm$) in the directions of the arrows. To obtain the correct matrix elements for the configurations related by symmetry to the ones shown, the hopping directions have to be transformed accordingly. Bold \textbf{1}s indicate leapfrog contributions. Note that leapfrog can take place regardless of the configuration in the neighboring plaquettes, and that all configurations for which leapfrog is allowed are connected to precisely 2 other configurations.}
\label{tab:hoppings-confined}%
\end{table}

The crudest way to account for the overall quantum motion of a dipole is to construct a tight-binding model with averaged hoppings. There are 5 elementary dipole configurations, which give rise to 20 total dipole states by rotations and / or reflections (see Table~\ref{tab:hoppings-confined}). If we define the dipole position as that of the common spin between the two monopoles, we see that the dipole (or bi-spinon) hops on a checkerboard lattice --- see Figs.~\ref{fig:leapfrog} and~\ref{fig:noleapfrog}. From the local spin configuration we determine the effective dipole hopping parameters as follows: (i) For each of the elementary dipole states, which are weighted by the probability of finding them in a given cluster, we count all neighboring-plaquette configurations that allow dipole hopping in a given direction. (ii) We divide the number we obtain in step (i) with the total number of configurations compatible with the dipole. And finally, (iii) after exhausting the dipole configurations, we take the average of the results of step (ii) to obtain the averaged matrix elements for each direction. Note that steps (i) and (ii) are performed approximately, by considering only spins in the immediate vicinity (i.e., nearest-neighboring plaquettes) of the dipole. 

The outcome of the averaging procedure is shown in Table~\ref{tab:hoppings-confined}: a tight-binding model for the dipole on a checkerboard-lattice with nearest and next-nearest hopping of magnitude $t_1 / J_\pm \simeq 0.8069$ and $t_2 / J_\pm \simeq 0.3257$. By forming a coherently propagating dipole described by this tight-binding model, the system with two excitations can lower its energy to $\sim - 3.88 J_\pm$. This energy is lower than the $-2\sqrt{2} J_\pm$ ($-2 J_\pm$) that can be reached via the reconnection (leapfrog) process. We shall use this effective model to compare to numerical results in the next Section. Note that the effective tight binding thus obtained contains no free parameters.

\subsubsection{Deconfined propagation}

The mechanism of lowering the total energy by forming coherently hopping quasiparticles is also expected to apply to deconfined spinons. It is therefore illustrative to compare the case in which spinons remain confined and behave as a single quasiparticle with the one where they propagate as individual particles. To do this, we now analyze the motion of independent spinons within the square ice background. We derive an effective tight-binding model with averaged hoppings, as done in Sec.~\ref{subsec:confined}, to describe the individual propagation of spinons. This treatment will allow us to anticipate that the system with two deconfined spinons can further lower its energy compared to the confined case.

When we remove the artificial confinement, spinons are allowed to move freely with respect to one another. Once again, we can derive an effective tight binding model for each of them following the same procedure as the one used for dipoles, except that in this case we only need to take into account a single ice rule-violating plaquette and its 4 immediate neighbors (see Table~\ref{tab:hoppings-deconfined}). Note that this approximate averaging ignores configurations with nearest-neighboring monopoles. Spin exchange enables the spinons to hop independently, each one on its own square sublattice, with average hoppings that are found to be simply $t_1 / J_\pm = 1$ and $t_2 / J_\pm = 0.5$. Again, there is no free parameter. 

\begin{table}[t]
  \centering
   \begin{tabular*}{\columnwidth}{ccccccccccc}
    \hline
    Configuration & Multiplicity & Amplitude & $\nearrow$ & $\searrow$ & $\swarrow$ & $\nwarrow$ & $\rightarrow$ & $\leftarrow$ & $\uparrow$ & $\downarrow$ \\
    \hline
     \smon & 4  & 0.25 & $\frac43$ & $\frac43$ & $\frac23$ & $\frac23$ & $\frac23$ & $\frac23$ & $\frac23$ & 0 \\
    \hline
  \end{tabular*}
  \caption{Monopole configuration, its multiplicity under rotations, amplitude in $a_{\bm{k}}^\dagger \ket{\mathrm{RK}}$, and cumulative hopping amplitudes (in units of $J_\pm$) in the directions of the arrows. To obtain the correct matrix elements for the configurations related by symmetry to the ones shown, the hopping directions have to be transformed accordingly.}
\label{tab:hoppings-deconfined}%
\end{table}

Since a single spin flip necessarily generates a monopole-antimonopole pair, one can make the creation and annihilation of the latter explicit by appropriately attaching monopole operators to the boson operators in Eq.~\eqref{eq:model}, i.e.,
\begin{equation}
 a_{\bm{i}} \;\; \rightarrow \;\; \tilde a_{\bm{i}} = a_{\bm{i}} \,  
 b_{\XBox}^\dagger b_{\XBox'}^{\ } \,, \label{eq:sub}
\end{equation}
where $\XBox, \XBox'$ are the two crossed plaquettes to which site $\bm{i}$ belongs. The hamiltonian ${\cal H}$ becomes
\begin{equation}
 {\cal H} = -J_\pm \sum_{\langle \bm{i},\bm{j} \rangle} ( a_{\bm{j}}^\dagger F_{\bm{ij}} a_{\bm{i}} 
 b_{\XBox}^\dagger b_{\XBox'} 
 b_{\XBox''}^\dagger b_{\XBox'''} 
 + \mathrm{h.c.} ) \,.
\end{equation}
The only hopping processes within the two-spinon sector are those for which $\XBox' = \XBox''$ or $\XBox = \XBox'''$, so that the term becomes quadratic in $b$ operators. Now, performing the averaging discussed above amounts to dropping the original $a$ operators and replacing the hamiltonian with the averaged tight binding. Explicitly, the monopole model is
\begin{eqnarray}
 {\cal H}_{\mathrm{monopole}} &=& - t_1 \sum_{\langle \XBox, \XBox' \rangle} ( b_{\XBox}^\dagger b_{\XBox'} + \mathrm{h.c.} ) 
 \nonumber \\ 
 && - t_2 \sum_{\langle\langle \XBox, \XBox' \rangle\rangle} ( b_{\XBox}^\dagger b_{\XBox'} + \mathrm{h.c.} ) \,,\label{eq:spinon}
\end{eqnarray}
where the shorthands $\langle \XBox, \XBox' \rangle$ and $\langle\langle \XBox, \XBox' \rangle\rangle$ now denote nearest and second-nearest neighbors on the monopole square lattice, respectively.

The spectral-function approach we used previously can be adapted for the fitting of the DSF in the deconfined case. In order to take both monopoles into account simultaneously, we need to generalize the spectral function, i.e., the density of states, to a two-particle energy-resolved joint density of states (EJDOS). This quantity is written as
\begin{equation}
 \tilde {\cal S} (\bm{q},\omega) = \sum_{\bm{k}} |\braket{\bm{q}-\bm{k};\bm{k}| b_{\bm{q}-\bm{k}}^\dagger b_{\bm{k}} |0}|^2 \delta(\omega - \epsilon_{\bm{q}-\bm{k};\bm{k}} ) \,,\label{eq:ejdos}
\end{equation}
where $\ket{\bm{q}-\bm{k};\bm{k}}$ is a two-spinon eigenstate of ${\cal H}_{\mathrm{monopole}}$, with the two spinons at momenta $\bm{q}-\bm{k}$ and $\bm{k}$ and corresponding eigenenergy $\epsilon_{\bm{q}-\bm{k};\bm{k}}$. Here, in the spirit of the averaging approximation, we have replaced the spinon vacuum, i.e., the RK state, with the true vacuum. The EJDOS generically has broad features even for free systems, so we anticipate that a one-to-one peak matching is unlikely. The goal of this comparison is to capture the dispersion of the bottom of the interacting two-spinon continuum. However, we can a priori say the following: the lowest-energy single-spinon state of the tight-binding model derived in this section is $-6 J_\pm$. Since a spin flip introduces two spinons to the system, this model predicts a minimum two-spinon energy of $-12 J_\pm$, much lower than the $\sim-3.88 J_\pm$ of the confined propagation. From the effective treatments of this section we can thus anticipate that it is energetically favorable for two spinons in quantum square ice to propagate coherently and independently as deconfined particles. In the following Section, we compare the results obtained from the effective theories introduced in this Section to exact numerics on finite clusters.

\section{Results}\label{sec:results}

\begin{figure}[t]
\centering
\includegraphics[width=1.\columnwidth]{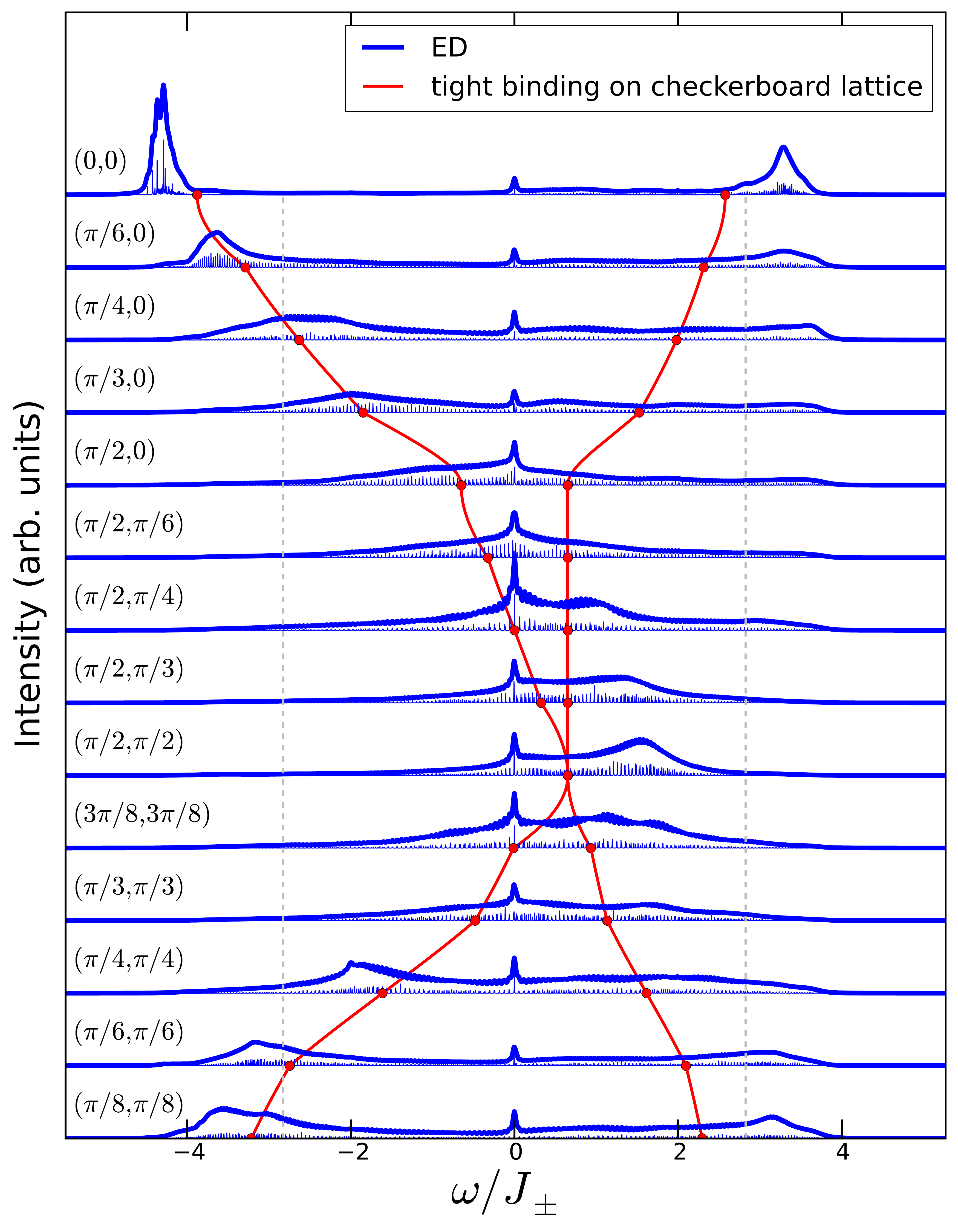}
\caption{Dynamic spin structure factor for the confined dynamics along the high-symmetry path in Fig.~\ref{fig:ckb}(b) evaluated by Lanczos exact diagonalization of 72- and 64-site clusters. For spectra at momenta that can be obtained from both clusters, the one from the 72-site cluster is shown. Thin blue lines represent bare peaks; thick blue lines include a Lorentzian broadening. Red lines and dots show the bands of the dipole checkerboard-lattice model derived in Sec.~\ref{sec:model} with $t_1 / J_\pm \simeq 0.8069$ and $t_2 / J_\pm \simeq 0.3257$. Dashed gray lines denote the energy regime of leapfrog and reconnection processes.}
 \label{fig:akw-conf}
\end{figure}

We have performed large-scale exact diagonalization calculations to test the effective models developed in the preceding section. We begin by discussing the dynamic structure factor for the confined case. Our results are summarized in Fig.~\ref{fig:akw-conf}. The DSF can be roughly separated into two energy regimes. The first is contained in the range $-2\sqrt{2} < \omega/J_\pm < 2\sqrt{2}$, which is the bandwidth of the reconnection process. In this regime, we expect coherent dipole formation to compete with leapfrog and reconnection. We see that the dominant zero-energy peak associated with propagation on Lieb chains survives, with an intensity modulation along $\bm{k}$ matching that of the DSF of ${\cal H}_{\mathrm{reconnection}}$. The second regime, outside the aforementioned band of energies, is where quasiparticle coherence is expected. Indeed, more distinctive features can be observed in this energy range, even though peaks are still considerably broad. The sharp, lowest-energy peak at the bottom of the band at $\bm{k}=0$ arises due to the fact that $a_{\bm{k}=0}^\dagger\ket{\mathrm{RK}}$ is very close to the ground state of ${\cal H}$. In fact, when $J_z$ is finite but much larger than $J_\pm$, the two become identical~\cite{Pollmann2006}.

\begin{figure}[t]
 \centering
 \includegraphics[width=1.\columnwidth]{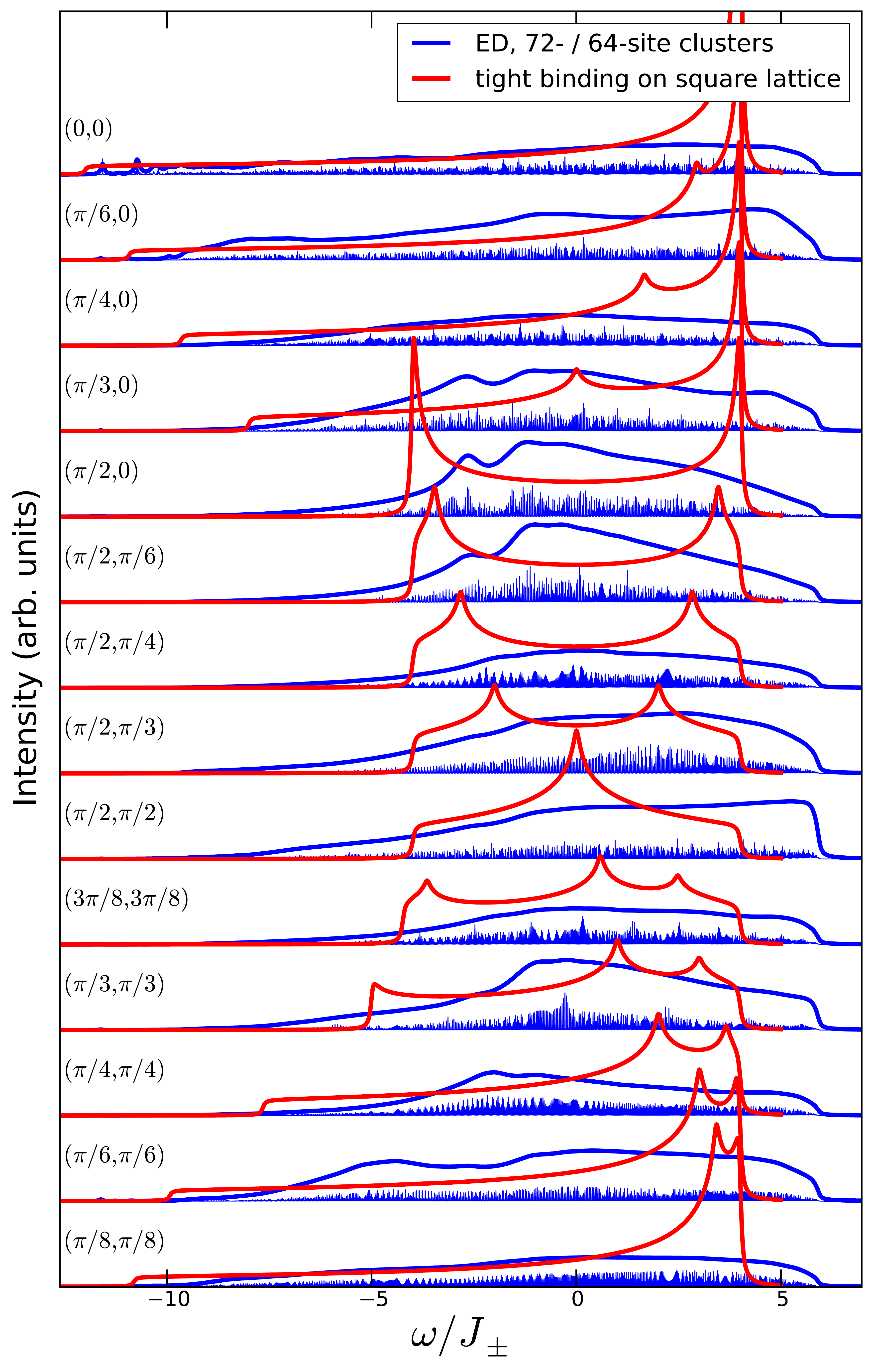}
 \caption{Dynamic spin structure factor along the high-symmetry path in Fig.~\ref{fig:ckb}(b) evaluated by Lanczos exact diagonalization of 72- and 64-site clusters. For spectra at momenta that can be obtained from both clusters, the one from the 72-site cluster is shown. Thin blue lines represent bare peaks; thick blue lines include a Lorentzian broadening. Red lines show the energy-resolved joint density of states for the monopole square-lattice model derived in Sec.~\ref{sec:model} with $t_1 / J_\pm = 1$ and $t_2 / J_\pm = 0.5$.}
 \label{fig:akw}
\end{figure}

Notwithstanding the simplicity of the effective dipole theory derived in the previous Section, we find that it follows the overall DSF rather well. The lower band, in particular, seems to follow the dominant finite-energy peak even within the energy range of the leapfrog and reconnection processes. The upper band follows the features in the DSF much less accurately. Discrepancy is expected due to several reasons. Firstly, the finite size of the cluster may preempt the development of fully coherent dipoles, which are rather extended objects compared to point-like particles. Secondly, the effective tight binding takes into account correlations with the spin background only locally and on average. A partial remedy to this might require to account for the interaction-related dressing in the confined dynamics. Finally, it seems like the averaging is overestimating the bandwidth suppression that is caused by the scattering of the dipole off of the spin background. These issues aside, the lower bound of the DSF is followed rather closely by our simple free theory, thus hinting at free delocalized dipoles.

We now turn to the deconfined dynamics, which is the most challenging case. We compute the EJDOS of Eq.~\eqref{eq:ejdos} for the effective spinon tight-binding model ${\cal H}_{\mathrm{monopole}}$ and compare it to the DSF in Fig.~\ref{fig:akw}. As we have anticipated, the EJDOS, albeit significantly broad, has several sharply defined features, reminiscent of van Hove singularities, that have no counterparts in the DSF of the interacting system. If the processes that give rise to the cusps in the EJDOS take place in the fully interacting model, then interference with other processes must be smearing them out. We do, however, observe a reasonable agreement at low energies, as the EJDOS accounts well for the lower bound of the DSF and follows its dispersion faithfully almost up to the BZ boundary. It is noteworthy that a very different approach applied to 3D quantum spin ice yields comparable results~\cite{Wan2015a}: well-defined quasiparticle behavior at the zone center, gradually fading towards the zone boundary. The lowest-energy peak in the DSF is again at $\bm{k}=0$ and is sharp due to the closeness of the underlying state to $a_{\bm{k}=0}^\dagger\ket{\mathrm{RK}}$. As in the confined case, the features at high energies are inaccessible to the noninteracting spectrum.

\section{Discussion and concluding remarks}\label{sec:conclusions}

In summary, we have derived an interacting model of spinon excitation dynamics in two-dimensional quantum spin ice from the strong-Ising limit of the checkerboard-lattice XXZ model. We have studied this model in two interesting regimes. When monopoles are artificially confined to nearest-neighbor distance, they propagate via a combination of the two mechanisms detailed in Sec.~\ref{subsec:confined}, which we understand in terms of self-avoiding paths (leapfrog) and Lieb chains (reconnection). The combination of the two processes allows the system to lower its energy by forming a coherently propagating monopole-antimonopole composite. Via an averaging procedure of possible hoppings within the square ice background, we derive an effective tight-binding model that describes the kinetics of this emergent quantum particle, leading to a concrete prediction for its energy dispersion. We compare this dispersion to that of the DSF of the original interacting model, including the artificial confinement constraint, and find compelling evidence of coherent free propagation of the composite excitation.

Having unraveled the energy-lowering mechanism of formation of coherent quasiparticles in the confined regime of excited quantum square ice, we then apply the same argumentation to the more challenging case of unconstrained spinon motion. Here we distinguish features that unequivocally signal the separate propagation of the two monopoles as free independent particles and contrast those of the confined dynamics. In particular, our results demonstrate that the lower bound of the two-spinon continuum in the numerically obtained DSF has a characteristic dispersion that is in good agreement with a free spinon theory, up to minor corrections due to interactions. We thus uncover fingerprints of coherent propagation of free spinons in quantum square ice.

This finding suggests that it is possible to obtain evidence of fractionalized coherent quasiparticles in the dynamic response of quantum spin ice, above the Ising energy scale. Assuming, for example, a value of $J_\pm \sim 0.05$~meV, which is presumably the relevant one for \YTO~\cite{Ross2011}, we see that this lower-bound dispersion spans a range of approximately $6J_\pm\sim$ 0.3~meV, well within the resolution of state-of-the-art experiments~\cite{Hallas2016}. Our approach can also be used to put to the test the interpretation proposed in recent spectroscopy and transport studies of coherently propagating monopoles in quantum spin ice materials~\cite{Pan2015,Tokiwa2015}. In order to make a direct connection to experiments, however, our methodology will have to be extended to three dimensions and nonzero temperatures. On the other hand, optical lattices of bosonic cold atoms can offer a viable platform to test our present results directly~\cite{Glaetzle2014}. It would also be interesting to see whether there is applicability to other two-dimensional quantum spin liquid candidates~\cite{Yamashita2010}. 

When drawing a parallel between our toy model and realistic hamiltonians for quantum spin ice, some remarks are in order. Firstly, compared to the prototypical model derived by fitting inelastic neutron scattering measurements for ferromagnetically ordered Yb${}_2$Ti${}_2$O${}_7$ in a field~\cite{Ross2011}, Eq.~\eqref{eq:xxz} is missing two couplings. The first one is usually denoted by $J_{\pm\pm}$ and corresponds to pair creation or annihilation in the boson language; it was found to be small and is typically ignored. The second coupling, $J_{z\pm}$, was shown to be detrimental to the quantum $U(1)$ liquid, favoring a splayed ferromagnetic phase instead~\cite{Hao2014}. Even though the term proportional to $J_{z\pm}$ is interesting in its own right and probably significant for modeling materials, it is not essential for the physics we investigate here. Secondly, the spinon dynamics in the system we investigate arises from double spin flips, caused by the transverse term already existent in the model. This contrasts and complements studies of spinon dynamics generated by an externally applied transverse field~\cite{Henry2014,Petrova2015,Wan2015a}. In a real material both processes will likely be present simultaneously. 

It should also be noted that spinon excitations were studied in a fermionic model similar to ${\cal H}$~\cite{Fulde2002}. The spectral function of that model was later calculated and was shown to contain characteristics of free fractional excitations~\cite{Pollmann2006}. Even though the model and its ground state are different in the two cases, the methodology followed is similar in spirit.

Our numerical calculations are performed in large systems for exact-diagonalization standards. Even so, we expect finite-size effects to be detrimental to long-range coherence. This is because a quasiparticle that is only approximately free may not have enough room to form fully within the area of the clusters, especially when it becomes effectively an extended object. Moreover, a finite cluster means that spinons are forced to be close to one another, which also exacerbates the side effects of (effective) spinon interactions. We therefore expect the tendency towards free coherent quasiparticle behavior to be even stronger in larger systems. It is compelling to find out whether there may be an energy threshold that defines a dynamic transition from free to correlated behavior.

\begin{acknowledgments}
The authors are grateful to B.~Dou\c{c}ot, D.~L.~Kovrizhin, R.~Moessner, J.~G.~Rau and Y.~Wan for helpful discussions. This work was supported in part by EPSRC Grant No.~EP/K028960/1, the EPSRC NetworkPlus on ``Emergence and Physics far from Equilibrium'', and Perimeter Institute for Theoretical Physics. Research at Perimeter Institute is supported by the Government of Canada through Industry Canada and by the Province of Ontario through the Ministry of Economic Development and Innovation. SK acknowledges financial support by the ICAM branch contributions. Part of the numerical calculations were performed using the Darwin Supercomputer of the University of Cambridge High Performance Computing Service, provided by Dell Inc.~using Strategic Research Infrastructure Funding from the Higher Education Funding Council for England and funding from the Science and Technology Facilities Council. 
Statement of compliance with EPSRC policy framework on research data: this publication reports theoretical work that does not require supporting research data.
\end{acknowledgments}

\appendix

\section{Methods}\label{sec:methods}

\begin{figure}[]
 \centering
 \includegraphics[width=\columnwidth]{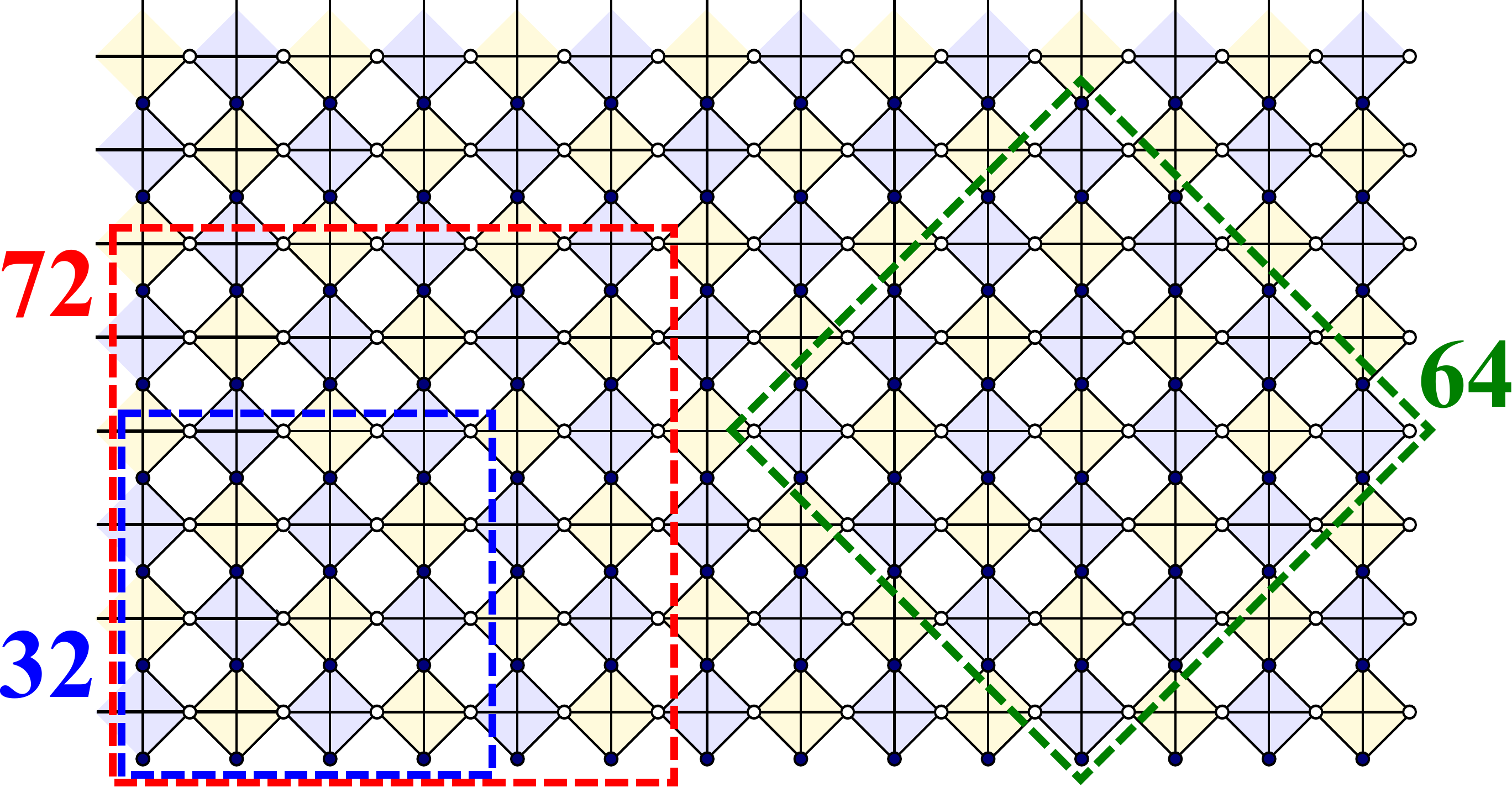}
 \caption{The clusters used in our exact-diagonalization calculations.}
 \label{fig:clusters}
\end{figure}

We use 3 different clusters in our numerical calculations, shown in Fig.~\ref{fig:clusters}. The diagonalization of the hamiltonian matrices we consider in this work is performed using the Lanczos method. A technicality that deserves commenting is the targeting of the relevant subspaces of the full Fock space. Clearly, the numerical treatment of a $2^L \times 2^L$ hamiltonian with $L=72$ would be impossible without restriction to the states (partially) obeying the ice-rule constraint, even with use of all space-group symmetries. We have used two different strategies to generate the manifold of states with two violations of the ice rule. The first is the one described in Sec.~II.B.2 of Ref.~\onlinecite{Zhang2003}: starting from an initial state, we generate the Fock space recursively by repeatedly applying the hamiltonian to the states of the previous step. The algorithm terminates when this procedure stops yielding new states. This method allows us to check for ergodicity by comparing the resulting basis to a naive trial-and-error search through the entire unconstrained Hilbert space. In lieu of a general proof, we have checked that ${\cal H}_{M_{+2}}$ is ergodic on all clusters of up to 32 sites. The second method is a two-step process of first enforcing the ice-rule constraint to obtain all compatible configurations in an elementary plaquette, and then stacking two copies of the elementary plaquette. The process is restarted with the resulting larger plaquette as elementary. Once we have reached the desired system size, we enforce periodic boundary conditions and translational invariance.

\section{Rokhsar-Kivelson point for excited quantum square ice}\label{sec:rk}

When an ice-rule violating plaquette is surrounded by perfect square ice, there are precisely 6 configurations that can be accessed by the action of ${\cal H}$ of Eq.~\ref{eq:model}. Picturing the Fock space of configurations as a set of nodes and the hamiltonian as a graph connecting the nodes, configurations in which the monopole and antimonopole are farther than nearest-neighbours are nodes of degree 12. On the other hand, configurations in which monopole and antimonopole share a majority spin correspond to nodes of degree 10. (Note that if the shared spin is majority for one monopole, it must be majority for the other as well, or the two monopoles would have the same charge, which is forbidden in the $M_{+2}$ sector. On the other hand, the shared spin could be minority for both monopoles, in which case the corresponding node has degree 12; this case corresponds to a non-contractible pair discussed in Ref.~\onlinecite{Castelnovo2010}). 

By introducing a diagonal term, written symbolically as
\begin{subequations}
\begin{align}
 {\cal H}_\mu = -2 \mu \sum_{\bm{i}\in\Lambda_A} & \left( \ket{{}_{\bm{i}}\tlbleq} \bra{{}_{\bm{i}}\tlbleq} + \ket{{}_{\bm{i}}\tlbreq} \bra{{}_{\bm{i}}\tlbreq} \right. \\
  & + \ket{{}_{\bm{i}}\tlbdeq} \bra{{}_{\bm{i}}\tlbdeq} + \ket{{}_{\bm{i}}\tubdeq} \bra{{}_{\bm{i}}\tubdeq} \\
  & + \left. \frac{\pi}{2}\textrm{-rotated and / or $x$/$y$-reflected} \right) \, ,
\end{align}
\label{eq:chempot}%
\end{subequations}
the state graph that represents the hamiltonian ${\cal H}+{\cal H}_\mu$ is regularized for $\mu=J_\pm$. In physical terms, this means that ${\cal H}+{\cal H}_\mu$ is at a RK point and therefore has a quantum $U(1)$ liquid ground state of energy $E_{\mathrm{RK}}/J_\pm = -12$. Reaching this RK point in the two-spinon sector requires fine tuning; however, the contribution in Eq.~\eqref{eq:chempot} to the ground state becomes statistically insignificant in the dilute-spinon limit~\cite{Wan2015a}. This is corroborated by quantum Monte Carlo results, which indicate that any finite density of excitations promotes the RK point of quantum square ice from a critical point to a full-fledged phase~\cite{Henry2014}.

\bibliographystyle{apsrev4-1}
% \bibliography{/home/sk/Documents/phys/bibtex/library}
% \bibliography{bib}


\end{document}